\documentclass{rspublic}

\usepackage{graphicx}
\usepackage{amsmath}
\usepackage{natbib}

\newcommand{\tr}{\ensuremath{\mathrm{tr}}}
\renewcommand{\rho}{\varrho}

\newcommand{\Tr}{\ensuremath{\mathrm{Tr}}}
\renewcommand{\Re}{\ensuremath{\mathrm{Re}\,}}

\begin{document}

\title[Non-Markovianity with initial correlations]{\bf Non-Markovianity: initial correlations and nonlinear optical measurements}
\author[Dijkstra and Tanimura]{Arend G. Dijkstra\footnote{dijkstra@kuchem.kyoto-u.ac.jp} and 
Yoshitaka Tanimura\footnote{tanimura@kuchem.kyoto-u.ac.jp}}
\affiliation{Department of Chemistry, Graduate School of Science, Kyoto University, Kyoto 606-8502, Japan}

\label{firstpage}
\maketitle
\begin{abstract}{non-Markovianity, initial correlations, nonlinear optics}
By extending the response function approach developed in nonlinear optics, we
analytically derive an expression for the non-Markovianity [Laine, et al, Phys.
Rev. A 81, 062115 (2010)] in the time evolution of a system in contact with a quantum mechanical bath, and find a close connection with the directly observable nonlinear optical response. The result indicates that memory in the bath induced fluctuations rather than in the dissipation causes non-Markovianity. Initial correlations between states of the system and the bath are shown to be essential for a correct understanding of the non-Markovianity.
\end{abstract}

\section{Introduction} \label{sec:introduction}
For macroscopic systems, the second law of thermodynamics prescribes ever-increasing entropy. In fact,
decreases of entropy are permitted on short time scales. When the dynamics of small quantum systems on short time scales is studied, the flow of information between the environment and the system can be important [\cite{Waldram.1985.book}]. In a microscopic theory, there are three major effects of the environment (or bath) on the system. The first two are dissipation, which removes excess energy from the system, and fluctuations, which supply energy. These two effects are related through the fluctuation dissipation theorem, which assures that the correct finite temperature equilibrium state is reached. The third one, which is less well known outside the fields of nonlinear optics and NMR, is the presence of entanglement between system and environment states. This third effect plays a major role if the system bath interaction is strong, or if the characteristic time scale of the noise induced by the environment is slow compared to typical system time scales. It is the origin of a rephasing signal in photon echo and NMR echo measurements. 

The dynamics of a quantum system in contact with a bath is described theoretically by deriving an equation of motion for the reduced density matrix, which includes only the system degrees of freedom. It is often assumed that the characteristic time scale of motion in the environment is much shorter than anything that happens in the system. This approximation is convenient, because it allows the derivation of a closed equation of motion for the reduced density matrix. In particular, no information about the history of the dynamics can be stored in the bath, and the equation of motion is local in time. The requirements that the trace of the density matrix must be preserved, and that the diagonal matrix elements must be positive in any basis then lead to a master equation in the Lindblad form [\cite{BPbook, Lidar.2001.cp.268.35}]. The dissipation operators that appear in this equation can be specified by modeling the environment and its interaction with the system. Because of the fast bath approximation, commonly used in association with secular approximations [\cite{Fassioli.2010.jcpl.1.2139}], the master equation approach does not properly include the fluctuations caused by the bath. On the other hand, stochastic Liouville equations, which do treat the fluctuations correctly, neglect the dissipation [\cite{Tanimura.2006.jpsj.75.082001}].

The restriction to an environment with much faster dynamics than the system breaks down for many systems studied in ultrafast nonlinear optics and NMR. In addition, it is not valid at low temperature, when a correct quantum mechanical description of the bath introduces additional time scales determined by the Matsubara frequencies. Obviously, the dynamics becomes more complex in this situation. Although strong system bath interaction can be included [\cite{Jang.2008.jcp.129.101104, Nazir.2009.prl.103.146404}], the key difference with the Lindblad formalism is the presence of memory. An environment which is not infinitely fast (compared to typical time scales in the system) can store information about the past. This information can subsequently flow back into the system, influencing the dynamics. Such memory effects can be included in master equations. These can, however, usually not describe a second key effect of a slower environment: the presence of correlated superpositions of the system and the environment in the initial state [\cite{Suarez.1992.jcp.97.5101}]. Because such correlations introduce a second source of memory, they cannot in general be ignored.

The idea that the flow of information from the environment back to the system can be used to quantify the extent of memory in a non-Markovian quantum process has been developed by \cite{Breuer.2009.prl.103.210401} and \cite{Laine.2010.pra.81.062115}. In a memoryless situation, two system states that are initially a certain distance apart, will only get closer during the time evolution. Therefore, when states are found that grow farther apart, the time evolution can be called non-Markovian. By introducing distance measures on the space of system states, these ideas can be made precise, resulting in a measure for non-Markovianity that depends only on system degrees of freedom.

Once such theoretical measures have been introduced, the question arises how they can be measured in experiment. In principle, quantum state tomography yields the complete quantum state of the system [\cite{Kuah.2007.pra.76.042113, Mohseni.2010.pra.81.032102}]. Once this measurement has been performed, any quantity that is a functional of the reduced density matrix can be calculated. However, this process is rather cumbersome and indirect. More straightforward methods to quantify concepts such as entanglement and non-Markovianity from experiments are welcome [\cite{Xu.2010.pra.81.044105, Cramer.2011.prl.106.020401}].

Experiments that can achieve this goal are found in the field of nonlinear optics. Observables such as the photon echo and two-dimensional optical spectroscopy depend on multiple time intervals, and are sensitive to memory effects that extend over several of these intervals [\cite{Mukamel.2009.acr.42.1207, Engel.2007.nature.446.782}]. 

In this paper, we show how correlations between the system and the environment are of critical importance for the non-Markovianity. In particular, for a simple model environment the dynamics is completely Markovian if the initial state does not include such correlations, while it becomes non-Markovian if correlations are allowed to be present. We describe the generation of initial correlations during a preparation time and discuss the close connection with the nonlinear optical response.

\section{Trace distance and non-Markovianity}
For a classical stochastic process, the meaning of "Markovian" is clear [\cite{VanKampen.1981.book}]: the future depends only on the present state, and not on the past. In the case of a Gaussian process, its correlation function must be exponential to have Markovianity. In the quantum case, we will define Markovianity following \cite{Laine.2010.pra.81.062115}, although other approaches have been proposed as well [\cite{Rivas.2010.prl.105.050403}]. As explained in the Introduction, the definition is based on the distance between a pair of quantum states.

A convenient measure is given by the trace distance $D$ between two density matrices $\rho^A$ and $\rho^B$, which is defined as
\begin{equation}
  D(\rho^A, \rho^B) = \frac{1}{2} \Tr_S \sqrt{(\rho^A - \rho^B)^2},
\end{equation}
or half the sum of the square root of the eigenvalues of $(\rho^A - \rho^B)^2$. The subscript S in the trace indicates that it is taken over system degrees of freedom, in contrast with trace operations over the environment which we will encounter later. 

In an ergodic system, any initial state will evolve in time until it reaches a single well-defined equilibrium. If there is no memory in the bath, the dynamics can only bring the system closer to equilibrium. Because no information can flow from the environment to the system, the distance between a pair of initial states will decrease with time. This is the case in memory-less approaches such as the Lindblad master equation. Memory in the bath means that the bath stores information about the system at a previous point in time, which affects the dynamics. The extra information opens the possibility of temporary time evolution in the unnatural direction. This suggest that non-Markovianity can be measured by studying how much two states move away from each other. In the definition given by \cite{Laine.2010.pra.81.062115}, this quantity is studied by defining the change in the trace distance
\begin{equation}
 \sigma(t; \rho^A(0), \rho^B(0)) = \frac{\mathrm d}{\mathrm dt} D(\rho^A(t), \rho^B(t)),
\end{equation}
which is integrated over time to define the non-Markovianity
\begin{equation}
  \mathcal N(G) = \mathrm{max} \int_{\sigma > 0} \mathrm{d}t \sigma(t; \rho^A(0), \rho^B(0)).
\end{equation}
The maximum is taken over all combinations of initial states $\rho^A(0)$ and $\rho^B(0)$. This measure is based on the distinguishability of the two trajectories. 

Although the trace distance can be defined for two density matrices of any shape, it takes a particularly simple form for a two-level system. In this case, the matrix elements of a density matrix are written in a given basis as
\begin{equation}
  \rho = \left( \begin{array}{cc} \rho_{11} & \rho_{12} \\ \rho_{12}^* & 1-\rho_{11} \end{array} \right).
\end{equation}
The trace distance between $\rho^A$ and $\rho^B$ is found as
\begin{equation} \label{trd2ls}
  D(\rho^A, \rho^B) = \sqrt{(\rho^A_{11} - \rho^B_{11})^2 + |\rho^A_{12} - \rho^B_{12}|^2}.
\end{equation}

\subsection{Initial correlations}
Although the (statistical) state of the system's degrees of freedom at any point in time is completely described by the reduced density matrix, this is nevertheless not the complete story. During the time evolution, the system gets entangled with the environment [\cite{Lopez.2008.prl.101.080503}]. If the time scale of the environment is not very short, this will influence the state of the system at later points in time. Therefore, the presence of classical or quantum mechanical correlations between system and environment affects dynamic measures like the non-Markovianity, although they are not explicitly present in the reduced description. Although correlations are automatically produced during the time evolution, leading to non-Markovianity, they can also be present in the initial state. These initial correlations contribute to the non-Markovianity as well and should be included in a proper description. To see this explicitly, we denote the complete density matrix including all system as well as bath degrees of freedom as $R(t)$. Its matrix elements in the system subspace are still operators on the bath degrees of freedom. Because the system and bath taken together form a normal quantum system, the complete density matrix evolves coherently in time, as dictated by the complete Hamiltonian $H$. We can define a propagator $G$ which propagates the density matrix as $R(t) = G(t - t_0) R(t_0)$, which is given by $G(t - t_0) = \exp(-i H^\times (t-t_0)/\hbar)$. The notation $H^\times A = [H, A]$ denotes the commutator. The reduced density matrix, which operates only on the Hilbert space of the system, is found by taking the partial trace over the bath, $\rho(t) = \tr_B R(t)$. For a factorized initial state, $R(t_0) = \rho(t_0) R_B(t_0)$, the time evolution can be written as a dynamical map $\rho(t) = \Phi(t; t_0) \rho(t_0)$. 

However, in the case of a slow environment, it is not clear why initial correlations between system and bath states can be neglected and the factorization assumption may break down. This means that the complete density matrix cannot be written in the form $R(0) = \rho(0) R_B(0)$, where $R_B(0)$ is a density matrix in the Hilbert space of the bath. Instead, each matrix element in the system space may depend on the bath in its own way. The difference between uncorrelated and fully correlated equilibrium density matrices, which are given by $\exp(-\beta H) / \Tr \exp(-\beta H)$, has been studied recently by \cite{Smirne.2010.pra.82.062114} and is readily observable in the optical response [\cite{Uchiyama.2010.pra.82.044104}].

To study the effect of more general initial correlations, we introduce a preparation time. This method allows us to interpolate between an uncorrelated state and the properly correlated equilibrium. An initially uncorrelated state is allowed to evolve for a time $t_1$, during which correlations are formed. The dynamics of the thus obtained correlated state at time zero is then followed during a time $t_2$. For $t_1 = 0$, correlations between system and bath are absent, while for a long enough preparation time maximum correlation is reached.

\section{Model}
The coherent time evolution of the system is given by a Hamiltonian $H_\mathrm{S}$. We employ a commonly used model for the environment that includes the complete quantum mechanical behaviour of bath modes, yet is flexible enough to be solved to a certain degree. In this model, the bath modes are harmonic oscillators, which couple linearly to the system. The Hamiltonian for the bath and its coupling to the system is given by
\begin{equation}
  H_\mathrm{B} + H_\mathrm{SB} = \sum_\alpha \left(\frac{p_\alpha^2}{2 m_\alpha}  + \frac{1}{2} m_\alpha \omega_\alpha^2 x_\alpha^2 \right) - \sum_\alpha g_\alpha x_\alpha V.
\end{equation}
Here, $\alpha$ indexes the bath modes, which have coordinates $x_\alpha$, momenta $p_\alpha$ and masses $m_\alpha$. $V$ denotes any operator on the Hilbert space of the system, which couples to the bath modes with strength $g_\alpha$. All necessary information about the system bath interaction is contained in the spectral density $J(\omega) = \frac{\pi}{2} \sum_\alpha \frac{g_\alpha^2}{m_\alpha \omega_\alpha} \delta(\omega - \omega_\alpha)$ and the temperature $T$. The correlation function can be written as the inverse Fourier transform of the spectral density as
\begin{equation} \label{cf}
L(t) = \frac{1}{\pi} \int_{0}^\infty  \mathrm{d}\omega J(\omega) \left( \coth \beta \hbar \omega/2 \cos \omega t - i \sin \omega t\right). 
\end{equation}
Its real part corresponds to the fluctuations, which are a function of the inverse temperature $\beta = 1/k_B T$ ($k_B$ is the Boltzmann constant), whereas the imaginary part is the dissipation.
Because the statistics for linear coupling to a harmonic bath is the same as for a Gaussian process, multi-point correlations functions are redundant (they can be evaluated using Wick's theorem), and are not necessary for the calculation of the propagator. In the classical limit, the bath can be modeled by a stochastic process. The time evolution is then given by a stochastic Liouville equation, which can include initial correlations [\cite{Ban.2010.pra.82.022111}] and is suitable for the calculation of the nonlinear response [\cite{Jansen.2004.jcp.121.10577}].

To simplify the analytical treatment, we will describe the situation where the system-bath interaction commutes with the system Hamiltonian, $[V, H_S] = 0$, such that the exact dynamics becomes second order in the system-bath interaction [\cite{Ishizaki.2008.chemphys.347.185, Nan.2009.jcp.130.134106, Ban.2010.pla.374.2324}]. For an overdamped Brownian oscillator spectral density, the non-commuting case can be handled efficiently using the hierarchy of equations of motion approach [\cite{Tanimura.2006.jpsj.75.082001, Shi.2009.jcp.130.164518}].

Although the system Hamiltonian can be chosen freely, we will for definiteness focus on a two-level system. In the basis of its eigenstates, the system Hamiltonian is diagonal, with matrix elements $0$ and $\epsilon$. The system-bath interaction causes dephasing in the excited state, and $H_\mathrm{SB}$ has matrix elements $0$ and $\delta\epsilon(X)$, where the fact that $\delta\epsilon$ is an operator on the bath degrees of freedom is indicated explicitly by the notation $(X)$. 

In a linear response experiment, the system is brought out of equilibrium by an external pulse, and the subsequent time evolution is probed. Non-Markovianity during the evolution time can occur in two ways. Firstly, it can be caused by memory in the system bath interaction during the evolution time. A second source of non-Markovianity are initial correlations between the system and the environment, which are present at the time the impulsive force interacts with the system. Such correlations can be studied in detail using nonlinear experiments, involving multiple pulses. 

\section{Results}

\subsection{Trace distance as a function of a single time}

When a system is initially in a factorized state, the only source of non-Markovianity is the buildup of system bath correlations during the time evolution. Suppose that two initial density matrices are given by $R^A(0)$ and $R^B(0)$. We make the usual assumption (which we want to relax later) that the system can be separated from the bath, and that the bath is in thermal equilibrium. The complete density matrix is then written as the direct product of a system part and a bath part, $R^A(0) = \rho^A(0) \exp(-\beta H_B) / \Tr_B \exp(-\beta H_B)$, where the reduced density matrix is $\rho^A(0) = \Tr_B R^A(0)$. Similar relations are written for $R^B(0)$. Such factorized initial conditions are typically found in electronic resonant spectroscopy, where the thermal energy is much smaller than the electronic excitation energy. The equilibrium density matrix then only contains population in the ground state, given by $\rho_\mathrm{eq} = |1\rangle\langle1|$. From this state, one can create any factorized initial state by applying an impulsive external interaction $\rho^{A/B}(0) = U \rho_\mathrm{eq}$, where $U$ denotes a Liouville operator. 

The time evolution of the complete density matrix is given by coherent evolution 
\begin{equation}
R^A(t) = \exp(-i H^\times t/\hbar) R^A(0) = \exp(-i(H_S^\times + H_B^\times + H_{SB}^\times)t/\hbar) R^A(0).
\end{equation}
We assume that the system Hamiltonian commutes with the system-bath interaction. In the interaction picture with respect to the bath Hamiltonian, the time evolution of the reduced density matrix then becomes
\begin{equation}
\rho^A(t) = \exp(-i H_S^\times t/\hbar) \langle \exp_+(-\frac{i}{\hbar} \int_0^t \mathrm{d}\tau H_{SB}^{\times I}(\tau)) \rangle \rho^A(0),
\end{equation}
where $\langle \cdots \rangle = \Tr_B \cdots \exp(-\beta H_B) / \Tr_B \exp(-\beta H_B)$ and $\exp_+$ denotes the time ordered exponential. 
The evaluation of the average over the bath is a standard calculation, giving
\begin{equation}
\langle \exp_+(-\frac{i}{\hbar} \int_0^t \mathrm{d}\tau H_{SB}^{\times I}(\tau)) \rangle = \exp(-\frac{1}{\hbar^2} \int_0^t \mathrm{d}t' \int_0^{t'} \mathrm{d}t'' \langle H_{SB}^{\times I}(t') H_{SB}^{\times I}(t'')\rangle),
\end{equation}
which for the coherences reduces to $\exp(-g(t))$, while the populations are constant in time. The dephasing function is given by
\begin{equation} \label{gt}
g(t) = \frac{1}{\hbar^2} \int_0^t \mathrm{d}t' \int_0^{t'} \mathrm{d}t'' L(t''),
\end{equation}
with the correlation function $L(t)$ given by Fourier transforms of the spectral density according to equation \ref{cf} [\cite{Mukamel.1995.book}].

For a two-level system, the trace distance between two density matrices which are initially prepared as $\rho^A(0)$ and $\rho^B(0)$ can be readily evaluated using equation \ref{trd2ls}. It is found to be
\begin{equation}
  D(\rho^A(t), \rho^B(t)) = \sqrt{(\rho^A_{11}(0) - \rho^B_{11}(0))^2 + |\rho^A_{12}(0) - \rho^B_{12}(0)|^2 \exp(-2 \Re g(t))}.
\end{equation}
Because we study the pure dephasing case, the populations $\rho^A_{11}$ and $\rho^B_{11}$ are constant in time, while the coherences evolve according to the dephasing function $g(t)$. While the dephasing function contains an imaginary (dissipative) part, which causes a time-dependent shift in the effective frequency, only the real part appears in the trace distance. 
If two state are prepared at time zero with initial populations, the trace distance simplifies to
\begin{equation}
D(\rho^A(t), \rho^B(t)) = D(\rho^A(0), \rho^B(0)) \exp(-\Re g(t)).
\end{equation}
In this case, the trace distance is directly related to the dephasing function. It is important to notice that the non-Markovianity only depends on the fluctuation part of the bath contribution, represented by the real part of $g(t)$. Thus, one cannot reveal this effect
from Lindblad-like quantum master equations, which only includes the dissipative
part of bath contribution properly. However, stochastic Liouville equations may be useful for the study of non-Markovianity.

We are now in a position to analyse the conditions for which the dynamics is non-Markovian. 
According to the definition of the non-Markovianity, the dynamics is non-Markovian only if the trace distance between two density matrices increases with time. The time derivative of the trace distance is found to be $\dot D(t) = - \Re \dot g(t) \exp(-\Re g(t))$. Because the exponential of a real number is always positive, the time derivative can be positive only if $\Re \dot g(t) < 0$. From the definition in equation \ref{gt}, the time derivative is $\dot g(t) = \int_0^t \mathrm{d}\tau L(\tau) / \hbar^2$. We see that the trace distance can only increase if the real part of the correlation function is negative, and sufficiently negative. Although the relation between the trace distance and the dephasing function is more complex in the general case where $\rho^A_{11} \neq \rho^B_{11}$, the populations do not influence the question whether the dynamics is Markovian. Non-Markovian time evolution is found if the trace distance increases at a certain point in time. Because the trace distance is a positive quantity, its derivative is given by a positive constant times $-\Re \dot g(t)$ also in the case of different populations. The previous analysis therefore applies, even though the value of the non-Markovianity will be different. 

However, because we started from initial states where the system and bath are factorized, this treatment does not include initial correlations between the system and the bath. To study their effect, we next study the non-Markovianity after an initial preparation time. 

\subsection{Trace distance as a function of two times}
\begin{figure}
 \includegraphics[width=8cm]{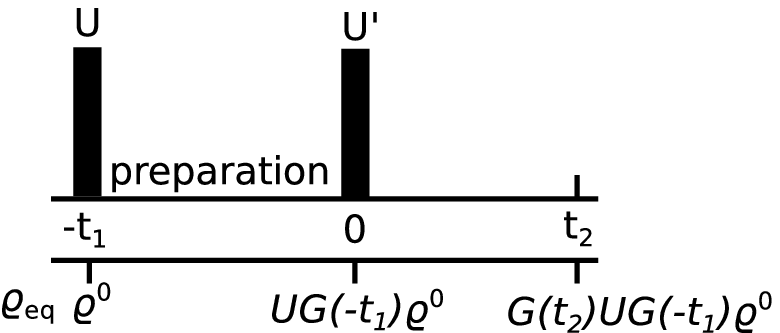}
 \caption{Schematic drawing indicating the time variables and the state of the reduced density matrix.\label{fig:schema}}
\end{figure}

To include initial correlations, we consider a preparation time. Starting from an state that factorizes into system and bath parts, which can be created as $\rho^0 = U \rho_\mathrm{eq} = U|1\rangle\langle 1|$ in optical experiments, the sample evolves during a time $t_1$. During this time correlations between the system and the bath form. An impulsive external force $U'$ is then applied to the system, after which time evolution takes place during an interval $t_2$. The time variables are illustrated in figure \ref{fig:schema}. The non-Markovianity during the time $t_2$ can now be caused by two effects: correlations that build up during $t_2$, as well as initial correlations present at the moment the external force interacts with the system, which are the result of the preparation.

The density matrix after evolution during two times is given by $\rho(t_1, t_2) = \Tr_B G(t_2) U' G(-t_1) \rho^0$, where we assume factorized conditions at time $-t_1$. $U'$ denotes a Liouville operator that models the second impulsive external force. The matrix product in the system Liouville space can be worked out explicitly by choosing a basis. We order the basis states as $|1\rangle\langle 1|$, $|1\rangle\langle 2|$, $|2\rangle \langle 1|$, $|2\rangle \langle 2|$ and denote the matrix elements of $U'$ in this basis as $U'_{ij,kl}$. Assuming that $\left[ H_S, H_{SB} \right] = 0$, we find
\begin{equation} \label{prop}
  \rho(t_1, t_2) = \Tr_B \left( \begin{array}{llll}
     U'_{11,11} & \zeta_1^* U'_{11,12} & \zeta_1 U'_{11,21} & U'_{11,22} \\
     \zeta_2^* U'_{12,11} & \zeta_2^* \zeta_1^* U'_{12,12} & \zeta_2^* \zeta_1 U'_{12,21} & \zeta_2^* U'_{12,22} \\
     \zeta_2 U'_{21,11} & \zeta_2 \zeta_1^* U'_{21,12} & \zeta_2 \zeta_1 U'_{21,21} & \zeta_2 U'_{21,22} \\
     U'_{22,11} & \zeta_1^* U'_{22,12} & \zeta_1 U'_{21,21} & U'_{21,22} 
       \end{array} \right) \rho^0,
\end{equation}  
where 
\begin{eqnarray}
\zeta_1 &=& \exp(-i \epsilon t_1/\hbar) \exp_+(-\frac{i}{\hbar} \int_{-t_1}^0 \mathrm{d}\tau \delta\epsilon(X(\tau))), \nonumber \\
\zeta_2 &=& \exp(-i \epsilon t_2/\hbar) \exp_+(-\frac{i}{\hbar} \int_{0}^{t_2} \mathrm{d}\tau \delta\epsilon(X(\tau)))
\end{eqnarray}
are still operators on the bath degrees of freedom. 

The trace over the bath degrees of freedom can now be calculated analytically, using cumulant expansion or path integral methods. It results in dephasing functions, which depend only on a single time when the average over either $\zeta_1$ or $\zeta_2$ is taken, but explicitly on both times for the average of products of two $\zeta$ functions. Because we are interested in the effect of initial correlations, these terms, which cannot be factorized into separate contributions depending on $t_1$ and $t_2$ only, are the most relevant to our treatment. They contain the effect of memory that extends over the externally applied force. From equation \ref{prop}, we see that these interesting terms multiply the matrix elements of the external force that operate on the coherences $\rho_{12}$ and $\rho_{21}$. There are four such terms, two which leave the coherence unchanged, and two which interchange the two coherences. To focus clearly on the effect of initial correlations, we choose an operation that flips the coherence, while leaving the populations unaffected. Such a force is given by a Liouville operator with matrix elements $U'_{11,11} = U'_{22,22} = U'_{12,21} = U'_{21,12} = 1$, and all other elements zero. Writing out the matrix elements in equation \ref{prop}, the density matrix is then given by 
\begin{equation}
  \rho(t_1, t_2) = (\rho^0_{11}, \zeta_2^* \zeta_1 \rho^0_{21}, \zeta_2 \zeta_1^* \rho^0_{12}, \rho^0_{22}).
\end{equation}

Starting from two density matrices $\rho^A(0)$ and $\rho^B(0)$, with equal initial populations, the trace distance between them evolves in time as $D(\rho^A(t_2), \rho^B(t_2)) = D(\rho^A(-t_1), \rho^B(-t_1)) |\zeta_2^* \zeta_1|$. As in the previous case of a single time interval, the restriction to equal populations only changes the value of the trace distance, but not the question whether the dynamics is Markovian. Using the cumulant expansion, we find 
\begin{equation}
  |\zeta_2^* \zeta_1| = \exp\left[-2 \Re g(t_1) - 2 \Re g(t_2) + \Re g(t_1 + t_2)\right]. 
\end{equation}
This expression enables the straightforward evaluation of the trace distance and the non-Markovianity for any spectral density. The term $g(t_1 + t_2)$ indicates the effect of initial correlations present at the time of interaction with the impulsive force $U'$. Such correlations, which extend across the the excitation, cannot be treated by the conventional reduced equation of motion approach, which includes Redfield and Lindblad equations. This has been pointed out in the calculations of nonlinear optical observables by \cite{Ishizaki.2008.chemphys.347.185}.

\subsection{Two-level system with overdamped bath}
As a simple example, which allows a more detailed analytical treatment, we will discuss the case of an overdamped Brownian oscillator. The spectral density is given by $J(\omega) = 2 \eta \omega \gamma / (\omega^2 + \gamma^2)$, 
which gives the dephasing function [\cite{Tanimura.1990.pra.41.6676, Weiss.2008.book}]
\begin{eqnarray} \label{gtbo}
  g(t) &=& \frac{\eta}{\gamma} \cot(\hbar\beta\gamma/2)(\exp(-\gamma t) + \gamma t - 1) + \frac{4 \eta\gamma}{\hbar\beta}\sum_{n=1}^\infty \frac{\exp(-\nu_n t) + \nu_n t - 1}{\nu_n(\nu_n^2 - \gamma^2)} \nonumber \\
       &-& i (\eta / \gamma)(\exp(-\gamma t) + \gamma t - 1).
\end{eqnarray}
For this model, the imaginary (dissipative) part of the correlation and dephasing functions depends only on the single time scale $\gamma$. As can be seen from the first line of equation \ref{gtbo}, the real (fluctuation) part includes additional time scales dictated by the Matsubara frequencies $\nu_n = 2\pi n / \beta \hbar$. 
While the imaginary part of $g(t)$ becomes constant due to the Ohmic nature of
$J(\omega)$ for $\gamma \to \infty$, the real part is time dependent as long
as $\beta$ is small.  This indicates that the fluctuation part of the bath noise cannot be delta correlated, 
even if this approximation is valid for the dissipation part.
At high temperature \emph{compared to the time scale of the bath}, $\hbar\beta\gamma/2 \ll 1$, these quantum fluctuation terms can be dropped, and the dephasing function simplifies
\begin{equation}
  g(t) = (2 \eta / \beta \hbar \gamma^2)(\exp(-\gamma t) + \gamma t - 1) - i (\eta / \gamma)(\exp(-\gamma t) + \gamma t - 1).
\end{equation}
In the high temperature case given above, the time derivative of the dephasing function is clearly always positive, and, consequently, the non-Markovianity vanishes for a single time interval. 

\begin{figure}
 \includegraphics[width=8cm]{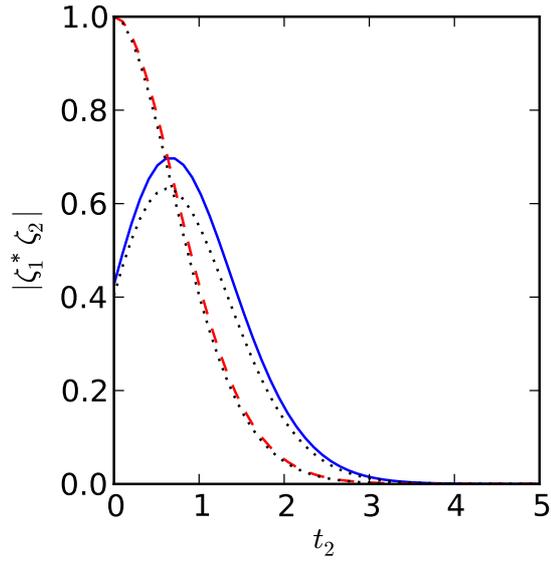}
 \caption{Trace distance as a function of time for preparation times $t_1 = 0$ (dashed line) and $t_1 = 1$ (solid line). The environment is modeled as an overdamped Brownian oscillator at high temperature with parameters $\beta \hbar \gamma = 0.5$ and $\beta \hbar \eta = 1.0$. The dotted lines show the same quantities calculated with 100 low temperature correction terms. Increase of the trace distance with time shows non-Markovian evolution. \label{fig:trd}}
\end{figure}

\begin{figure}
 \includegraphics[width=8cm]{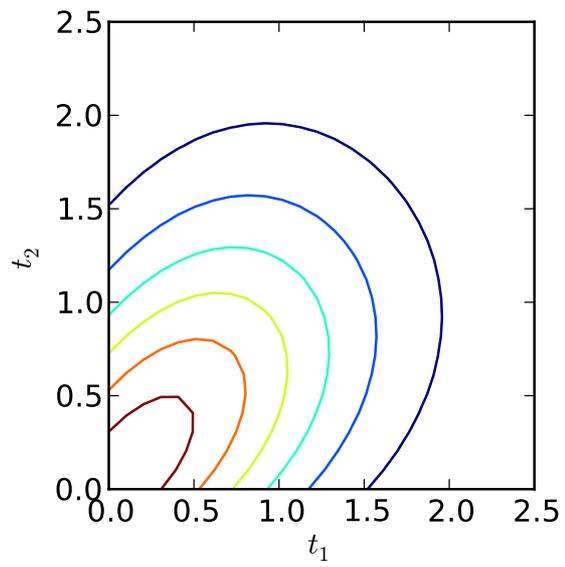}
 \caption{Trace distance as a function of two times, a preparation time $t_1$ and a detection time $t_2$. Parameters for the environment are the same as in figure \ref{fig:trd}. Correlations between the time evolution during $t_1$ and $t_2$ are clearly present. \label{fig:trd2t}}
\end{figure}

However, if we account for initial correlations by allowing them to form during a preparation time, the dynamics can become non-Markovian. This can be seen explicitly using the overdamped Brownian oscillator spectral density. For $t_1 = 0$, the trace distance varies with time as $\exp(-\Re g(t_2))$, and we recover the result found earlier. Because $\Re g(t) > 0$ for all times, the trace distance is strictly decreasing. Thus, the dynamics is Markovian for an exponential correlation function, in agreement with the classical definition. On the other hand, if we allow system bath correlations to form during the preparation time $t_1$, the trace distance can increase, and the measure for non-Markovianity is nonzero. This effect is shown in figure \ref{fig:trd}, where we compare a factorized initial state, corresponding to $t_1 = 0$, to a state that contains correlations, created by setting $t_1 \neq 0$. For a nonzero preparation time, the trace distance increases during a certain time interval, which shows that the dynamics is non-Markovian. It is clear that memory in the bath that extends over the pulse is crucial for this effect. The time evolution of the system after the application of the pulse is influenced by its state before the pulse, as can be seen from the presence of the $g(t_1 + t_2)$ term. For a bath that contains memory, this term does not factorize into functions of $t_1$ and $t_2$ only. The memory effect can be seen clearly by looking at the trace distance as a function of both times, as plotted in figure \ref{fig:trd2t}.

\subsection{Nonlinear optical response functions}
The time evolution of a quantum system during two intervals, separated by an external impulsive force, is closely related to nonlinear optical experiments. In these experiments, an initial pulse excites the system out of the ground state. Correlations between the system and the bath form during the following propagation time. After applying another pulse, the effect of these correlations can be observed.
As mentioned before, the initial state $\rho^0$ can be prepared by applying a pulse on the equilibrium state, $\rho^0 = U \rho_\mathrm{eq}$. In the case of optical experiments, the excitation energy is typically much larger than the thermal energy, and the equilibrium distribution only contains population in the ground state. The optical field couples to the dipole of the system, described by the dipole operator $\hat \mu = \vec \mu (|1\rangle \langle2| + |2\rangle \langle 1|)$. If we choose the operator $U$ as the commutator of the dipole with the density matrix, $U \rho = [\hat\mu, \rho]$, and furthermore set $U' = U^2$, the time evolution operators evaluated in the previous section correspond to the standard third-order nonlinear response functions with zero population time [\cite{Mukamel.1995.book, Tanimura.1996.jcp.106.2078}]. The observable in nonlinear optical experiments is the trace of the dipole operator multiplied with the density matrix, $\Tr_S \hat \mu \rho(t_1, t_2) = \Tr_S \hat \mu \Tr_B G(t_2) U' G(-t_1) U \rho_\mathrm{eq}$. By choosing the wave vectors of the incident pulses properly, it is possible to select pathways that are sensitive directly to the coherence flip described in the previous section. The resulting photon echo signal is given by [\cite{Nibbering.1991.prl.66.2464}]
\begin{equation}
  R(t) = \exp\left[-2 g(t_1) - 2 g(t_2) + g(t_1 + t_2)\right].
\end{equation}
Using heterodyne detection, both the real and imaginary parts of this response function are observable, while homodyne detection directly yields the absolute value. The connection with the non-Markovianity in the previous section is immediately clear: the photon echo is sensitive to exactly the memory effects that are responsible for non-Markovian dynamics. A photon echo experiment can be used to prepare a state in which the system and environment are correlated, and to subsequently probe the time evolution. Plotting the thus obtained response function directly answers the question whether the dynamics is Markovian or not, according to the definition given by \cite{Laine.2010.pra.81.062115}. 

\section{Conclusion} \label{sec:conclusion}
We have studied the non-Markovianity in quantum mechanical time evolution. This concept of Markovianity can be made precise by looking for states that become more distinguishable during time evolution. If such states are present, the process is clearly non-Markovian, which is the basic idea of the measure for non-Markovianity proposed by \cite{Laine.2010.pra.81.062115}. 
Non-Markovian time evolution corresponds to the presence of memory effects. Only the (temperature dependent) bath induced fluctuations, and not the dissipation, enter the non-Markovianity. We have treated the dynamics without making Markovian or rotating wave approximations, and thereby fully included the correlations between system and bath states, which influence the dynamics at a later point in time. Not only the correlations that are formed during the evolution, but also those present in the initial state can cause memory effects. By forming a correlated initial state during a preparation time, this effect can be studied for more general initial states than the equilibrium with respect to the complete Hamiltonian. We have shown that a process that is Markovian without initial correlations can become non-Markovian when such correlations are present. Conventional master equations, which cannot include the preservation of system bath entanglement across a pulse cannot be used to analyse this situation. Clearly, commonly used approximations such as a delta correlated bath or secular system-bath interaction don't hold either.
Because the procedure of preparing correlations during an initial time, and subsequently measuring their effect following an external impulse, the non-Markovianity is directly observable in nonlinear optical experiments such as the photon echo.
Future work should consider the three time photon echo, and the closely related two-dimensional optical spectra. In these experiments, population dynamics can be studied during a waiting time, allowing for more general measures of non-Markovian time evolution. Generalizations of the current work to more general system Hamiltonians, multiple baths, the case where the system Hamiltonian and the system-bath interaction do not commute and low temperature are possible using the hierarchy of equations of motion [\cite{Ishizaki.2005.jpsj.74.3131, Shi.2009.jcp.130.164518, Dijkstra.2010.prl.104.250401}].

\begin{acknowledgements}
AGD acknowledges the Japan society for the Promotion of Science for support in the form of a postdoctoral fellowship for foreign researchers. Y.T. was supported by a Grant-in-Aid for Scientific Research B2235006 from
the Japan Society for the Promotion of Science.

\end{acknowledgements}


\clearpage

\end{document}